\documentclass[prd,nopacs,nokeys,nofootinbib]{revtex4}
\usepackage{amsfonts}
\usepackage{amsmath}
\usepackage{amssymb}
\usepackage{bbm}
\usepackage[utf8]{inputenc}
\usepackage[caption=false]{subfig}
\usepackage{tensor}
\usepackage{slashed}
\usepackage[centertableaux]{ytableau}
\usepackage{hyperref}
\usepackage{natbib}
\usepackage{braket,mleftright}
\usepackage{graphicx}
\usepackage{cleveref}
\usepackage{tikz}
\usepackage{tikz,ifthen}
\usepackage{booktabs}
\usepackage{listings}
\usepackage{xcolor}
\begin{document}

\title{A Universal Convolution-Based Pre-processor to Correct the Prevalence-Incidence Gap in SIR, SEIR, and SIRS Modeling}

\author{Jos\'e de Jes\'us Bernal-Alvarado}
\email{bernal@ugto.mx}
\affiliation{Physics Engineering Department, Universidad de Guanajuato\\
Lomas del Bosque 103, Fraccionamiento Lomas del Campestre, 37150, Le\'on,
Guanajuato, M\'{e}xico}

\author{David Delepine}
\email{delepine@ugto.mx}
\affiliation{Physics Department, Universidad de Guanajuato\\
Lomas del Bosque 103, Fraccionamiento Lomas del Campestre, 37150, Le\'on,
Guanajuato, M\'{e}xico}


\date{\today}

\begin{abstract}
\textbf{Background:} Traditional compartmental models, including SIR, SEIR, and SIRS frameworks, remain the analytical standard for epidemic forecasting. However, real-world data validation consistently reveals significant predictive failures, such as peak underestimations of up to 50\%.

\textbf{Methods:} This research identifies a persistent "fundamental methodological error": the calibration of prevalence-based (stock) models using raw daily incidence (flow) data without proper transformation. We propose an integrated protocol utilizing an exponentially weighted convolution to reconstruct active cases from reported incidence:
\begin{equation}
I(t) \simeq \frac{1}{p} \int_{0}^{t} NDC(\tau) e^{-\gamma(t-\tau)} d\tau
\end{equation}
This transformation accounts for the recovery rate ($\gamma$) and the ascertainment rate ($p$).

\textbf{Results:} We demonstrate that increasing structural complexity, such as adding latency (SEIR) or waning immunity (SIRS), fails to resolve the incidence-prevalence gap. Simulation results show that without the proposed universal pre-processor, these advanced models inherit the systematic biases of misaligned data types. This leads to significant errors in estimating latent periods and the "heavy tail" of endemicity.

\textbf{Conclusions:} The proposed convolution transformation must serve as a universal prerequisite for any compartmental framework. This formal protocol bridges the gap between clinical reporting and mechanistic modeling. 


\end{abstract}

\maketitle
\section{Introduction}
Despite its widespread use in public health policy since 2019, the SIR model and its derivatives , such as the SEIR (Exposed) and SIRS (Waning Immunity) frameworks,\cite{kermack1927contribution,diekmann2000mathematical} consistently fails to capture real-world data trends due to their inherently rigid mathematical assumptions \cite{kermack1927contribution,diekmann2000mathematical,thompson2019modelling,moein2021inefficiency,nature2023validation,murph2025mapping}. In Real-world, their data validation consistently reveals significant predictive failures as most notably peak underestimations of up to 50\%\cite{murph2025mapping}.

A detailed revision of SIR model application allows to  identify a persistent "fundamental methodological error" in the calibration of these models: the use of raw daily incidence (flow) data to represent the infectious compartment, which is mathematically formulated as a prevalence (stock) variable. Usually, it is believed that adding compartments and increasing  the structural complexity of the models could compensate for data misalignment. But increasing structural complexity without proper data transformation merely propagates the incidence-prevalence gap into the transition rates of the new compartments (see Table \ref{tab:model_comparison}).

In this work, we propose a universal pre-processor data transformation that correct this data misalignment. A convolution formula is proposed to relate the New Daily Case (NDC) with the Infected $I(t)$ which appears in $SIR$-like model.  While the mathematical relationship between prevalence and incidence—modeled as a convolution—is well known (see for instance ref. \cite{diekmann2000mathematical}), a significant gap remains in its practical application. Existing literature often presents the convolution as a standalone proof rather than a universal requirement for compartmental calibration.

In section II, a review on $SIR$ model is briefly presented and in section III, the proposal to correct the prevalence-incidence methodological error is explained, applying it to simulated $SIR$ data. In this section, we propose two correction: the first one is enough to improve the precision on the peak position ($\mu$)of $I(t)$ and the second one is more general, it correct both the peak and also its amplitude, through a convolution product. In section IV, we explain that if the original data misalignment is not corrected, the generated error will propagate in $SIR$ extensions models as SEIR or SIRS models, generating errors in prediction on the behavior of the epidemic. In section V, the conclusion is presented.

\begin{table}[ht!]
\centering
\caption{Persistence of Methodological Risks across Compartmental Models}
\label{tab:model_comparison}
\begin{tabular}{|l|p{8cm}|}
\hline
\textbf{Model} & \textbf{Persistent Methodological Risk} \\ \hline
SIR  & Equating raw Incidence ($NDC$) with Prevalence induced errors in the peak localization of $I(t)$ and its amplitude (up to 50\% in some case). \\ \hline
SEIR & Generating errors in the estimate of the transition rate $S \to E$ and $E \to I$ by using raw daily reports. \\ \hline
SIRS  &Underestimating the ``heavy tail'' and misinterpreting reinfection rates. \\ \hline
\end{tabular}
\end{table}

\section{SIR model}
Let be the healthy people, susceptible to be infected (S), individuals with contagion capacity (I), and R representing the number of people recovered from the disease (or removed people because of death), dynamics of such populations, as a function of time, can now be described in the well known SIR model which  constitutes the foundational framework for mechanistic epidemiological modeling. It describes the flow of individuals between three mutually exclusive compartments governed by the following system of non-linear ordinary differential equations\cite{kermack1927contribution}:

\begin{equation}
\begin{aligned}
\frac{dS}{dt} &= -\beta \frac{SI}{N}, \\
\frac{dI}{dt} &= \beta \frac{SI}{N} - \gamma I, \\
\frac{dR}{dt} &= \gamma I,
\end{aligned}
\label{eq:SIR_system}
\end{equation}
where $N = S + I + R$ is the total population, $\beta$ represents the effective transmission rate, and $\gamma$ is the removal or recovery rate. It is important to note that in $SIR$ model, $I(t)$ also represents the active cases.  The fundamental epidemiological parameter, the Basic Reproduction Number ($R_0$), is derived from the stability analysis of the disease-free equilibrium and is defined as:

\begin{equation}
R_0 = \frac{\beta}{\gamma}.
\end{equation}

The dynamic evolution of the reproduction number as the susceptible compartment \(S(t)\) is depleted is known as the effective reproduction number, expressed as:\begin{equation} R_t = R_0 \frac{S(t)}{N} \end{equation}

While the SIR model successfully captures the initial exponential growth phase (where $S \approx N$), its deterministic structure imposes a strong symmetry on the incidence curve $I(t)$ around the peak. Specifically, the rate of decline is coupled to the depletion of the susceptible compartment. In modern pandemics modulated by social behavior, this symmetry is rarely observed; instead, the decline phase is often prolonged, resulting in a ``heavy tail" that the standard SIR model systematically underestimates.

We must remark that $I(t)$ describes the active population that is based on the daily incidence minus the recovered people.
\section{The Prevalence-Incidence Methodological Error}
A fundamental methodological error occurs when practitioners attempt to calibrate model parameters using daily incidence data (new cases), provided by organizations such as the WHO. The model is structurally formulated to utilize active cases (prevalence), a variable for which reliable statistics are rarely available.
In the traditional SIR model, New Daily Cases (NDC)—also referred to as Incidence—is expressed as the rate of flow of individuals from the Susceptible ($S$) compartment to the Infectious ($I$) compartment.
Mathematically, the NDC represents the negative rate of change of the susceptible population. In the continuous-time model, it is defined by the following term:

\begin{equation}
NDC(t) = -\frac{dS}{dt} = \beta \frac{S(t)I(t)}{N}
\end{equation}
\begin{itemize}
    \item $\beta$ (Transmission Rate): Represents the probability of infection per contact multiplied by the number of contacts per unit of time.
    \item $S(t)$: The number of individuals susceptible to the disease at time $t$.
    \item $I(t)$: The number of infectious individuals currently in the population (prevalence).
    \item $N$: The total population.
\end{itemize}
In real-world epidemiology, organizations like the WHO report data in discrete intervals (e.g., daily or every 5 days). To match this ``generated" clinical data, the NDC over a discrete time step $\Delta t$ is calculated as the total number of people who moved from $S$ to $I$ during that window:
\begin{equation}
    NDC(t+{\Delta t} )= S(t) - S(t + \Delta t) \approx \int_{t}^{t+\Delta t} \beta \frac{S(\tau)I(\tau)}{N} d\tau
\end{equation}
\subsection{The fundamental error}
\begin{figure}
    \centering
    \includegraphics[width=0.8\linewidth]{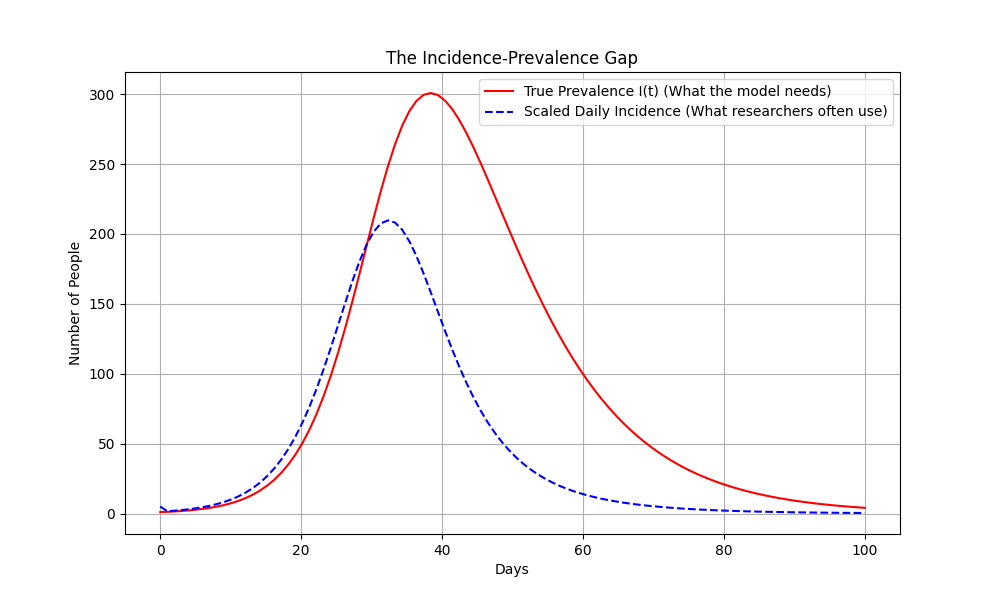}
    \caption{This figure illustrates the temporal ``gap" between disease prevalence ($I(t)$) and incidence within a standard SIR (Susceptible-Infectious-Recovered) model. The solid red line represents the true prevalence—the total number of infectious individuals at any given time—which determines the rate of new infections in the differential equations. The dashed blue line represents the scaled daily incidence, derived from the rate of change of the susceptible population ($-\frac{dS}{dt}$). The visualization highlights that incidence typically peaks earlier and declines faster than prevalence, a distinction that is vital for researchers when fitting models to daily case reports versus active clinical data.}
    \label{sir}
\end{figure}

A critical point raised in the provided analysis is that Incidence is not Prevalence.
\begin{itemize}
    \item $I(t)$ (Prevalence): Represents the state or the ``stock" of individuals currently ill.
    \item $NDC$ (Incidence): Represents the flow or the ``rate" of new arrivals into the ill state.
\end{itemize}
Equating the time derivative of cumulative cases directly with active cases $I(t)$ is identified as a fundamental methodological error that leads to inaccurate forecasting as it can be seen in figure(\ref{sir}).

The discrepancy arises because the model is structurally formulated to utilize active cases (prevalence), yet practitioners often attempt to calibrate it using daily incidence data (new cases).

\subsection{Mathematical Correction}
To bridge the gap between reported new daily cases ($NDC$) and the infectious compartment ($I$), we shall proceed in two step. First we shall propose a modification to compute $I(t)$ as a function of $NDC(t)$, this will permit to strongly improve the peak position and then we shall propose a mathematical proposal to extract $I(t)$ to fix also the amplitude of the peak.

\subsubsection{Peak Position correction}
To resolve this, we can define the relationship between the reported New Daily Cases (NDC) and the theoretical Infectious compartment ($I$).

\begin{figure}
    \centering
    \includegraphics[width=0.8\linewidth]{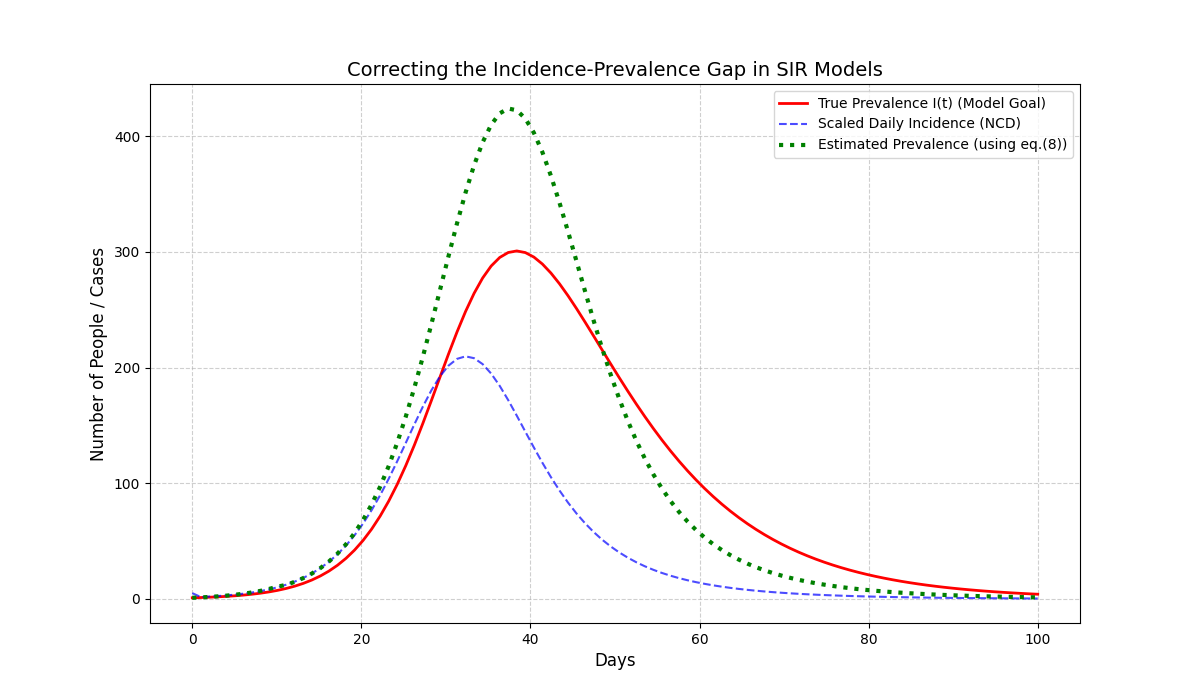}
    \caption{ The solid red line represents the  Prevalence ($I(t)$). The dashed blue line shows Scaled Daily Incidence (New Case Data), highlighting the characteristic ``peak offset" where incidence leads prevalence in time. The dotted green line shows the Estimated Prevalence, calculated by integrating new cases over a sliding window ($1/\gamma$) representing the average infectious period. This transformation successfully reconstructs the timing of the infectious pool.}
    \label{sir2}
\end{figure}

If we assume a discrete reporting interval $\Delta t$ (e.g., 1 day or 5 days), the relationship is\cite{diekmann2000mathematical}:
\begin{eqnarray}
I(t)&\simeq & \int_{t-1/\gamma}^t  NDC(\tau) d\tau \\
     I(t) &\simeq & \sum_{\tau=t-1/\gamma}^{t} NDC(\tau) 
\end{eqnarray}
Where $1/\gamma$ represents the average infectious period. The last line corresponds to the discretized case. This transformation accounts for the fact that prevalence is an accumulation of recent incidence minus those who have recovered. In figure(\ref{sir2}), one can see the effect of this correction.

\subsubsection{Peak amplitude correction}
As one can see from  figure(\ref{sir2}), the correction given above is inducing a large error on the amplitude of the peak. While the previous corrections  correctly shifts the timing of the peak, it often underestimates the magnitude because it treats the recovery process as a ``step function" (everyone stays infectious for exactly $1/\gamma$ days and then recover0s instantly). In reality, recovery is an exponential decay process.

To get the amplitude right, we need to weight the past incidence by the probability that those individuals are still infectious at time $t$. According to the SIR differential equations, the probability of remaining infectious after $\tau$ days is $e^{-\gamma \tau}$. 

Another problem is that the daily reported cases are far from the real number of infected people. Not every infection is captured in WHO datasets due to asymptomatic cases and testing delays. So one need not only to correct the peak amplitude taking into account that recovery is not a step function  but also the problem of data taking in a real world. 

To solve this last problem, we shall introduce an ascertainment rate $p$, where $0 < p \le 1$:
\begin{equation*}
    I_{reported}(t) = p \cdot I_{true}(t)
\end{equation*}
If $NDC(t)$ is the observed daily incidence, the true prevalence at time $t$ can be estimated by accounting for the recovery rate $\gamma$ by the following convolution product \cite{kermack1927contribution,diekmann2000mathematical,brookmeyer1988methods, keiding1991age}:
\begin{equation}
    I(t) \simeq \frac{1}{p} \int_{0}^{t} NDC(\tau) e^{-\gamma(t-\tau)} d\tau
\end{equation}
Where:
\begin{itemize}
    \item $NDC(\tau)$ (New Daily Cases): This is the incidence data provided by health organizations. It is the primary input generated from daily clinical reports.
    \item $p$ (Ascertainment Rate or Reporting Probability): This parameter accounts for data reporting biases. It represents the fraction of true infections actually captured by the healthcare system. It can be evaluated by comparing reported cases with seroprevalence surveys or any models to estimate under-reporting and asymptomatic volume.
    \item $\gamma$ (Recovery Rate): This defines the Infectious period ($1/\gamma$). In the model, it dictates how quickly individuals transition from the Infectious (I) to the Removed (R) compartment. It is derived from clinical data observing the average duration a patient sheds the virus or remains symptomatic.
    \item $e^{-\gamma(t-\tau)}$ (Survival Probability): we assume that the survival probability is following the universal exponential decay law. It weights past incidence by the probability that those individuals have not yet recovered by time $t$.
    \item $t - \tau$ (Time Lag): The time interval between the moment  when a case was originally reported ($\tau$) and the current time of evaluation ($t$)
\end{itemize}

\begin{figure}
    \centering
    \includegraphics[width=0.8\linewidth]{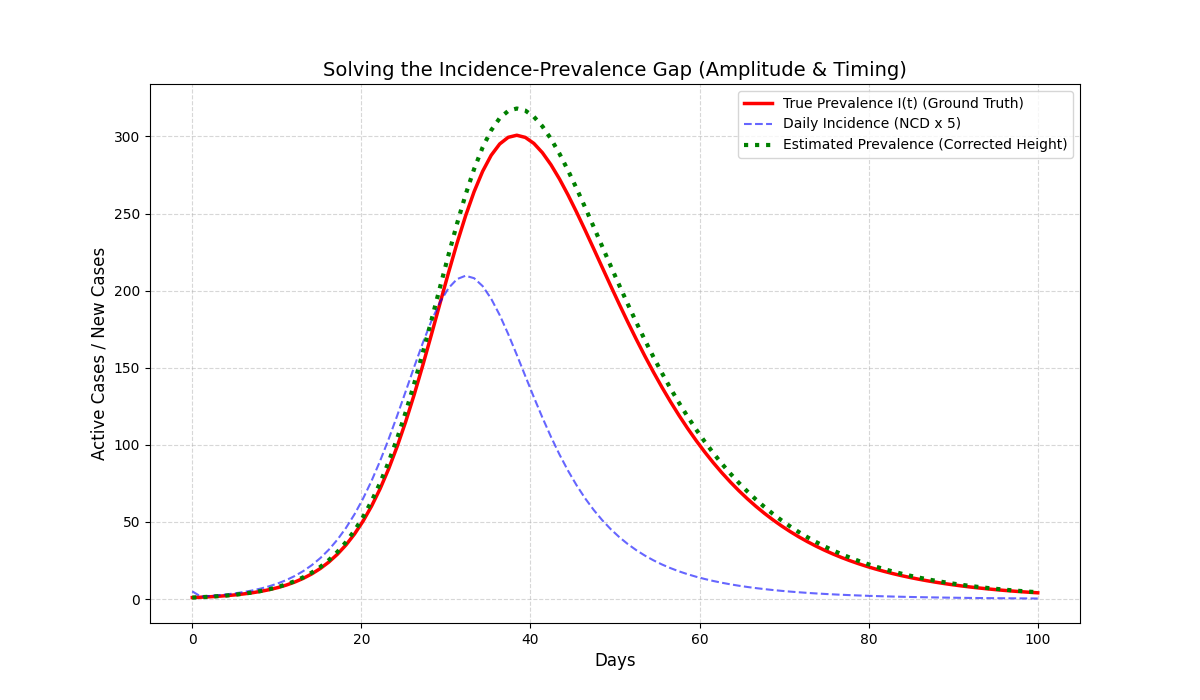}
    \caption{This figure shows the resolution of the ``Incidence-Prevalence Gap'' through an exponentially weighted transformation. The solid red line represents the True Prevalence ($I(t)$), the structural variable required for SIR model calibration. The dashed blue line depicts the Daily Incidence, which peaks prematurely and at a different magnitude compared to active cases. The dotted green line shows the Estimated Prevalence, reconstructed by weighting past incidence with an exponential survival function ($e^{-\gamma \tau}$)}
    \label{sir3}
\end{figure}

In figure(\ref{sir3}), one can see the results of the implementation of the correction given here. 

\begin{figure}
    \centering
    \includegraphics[width=0.5\linewidth]{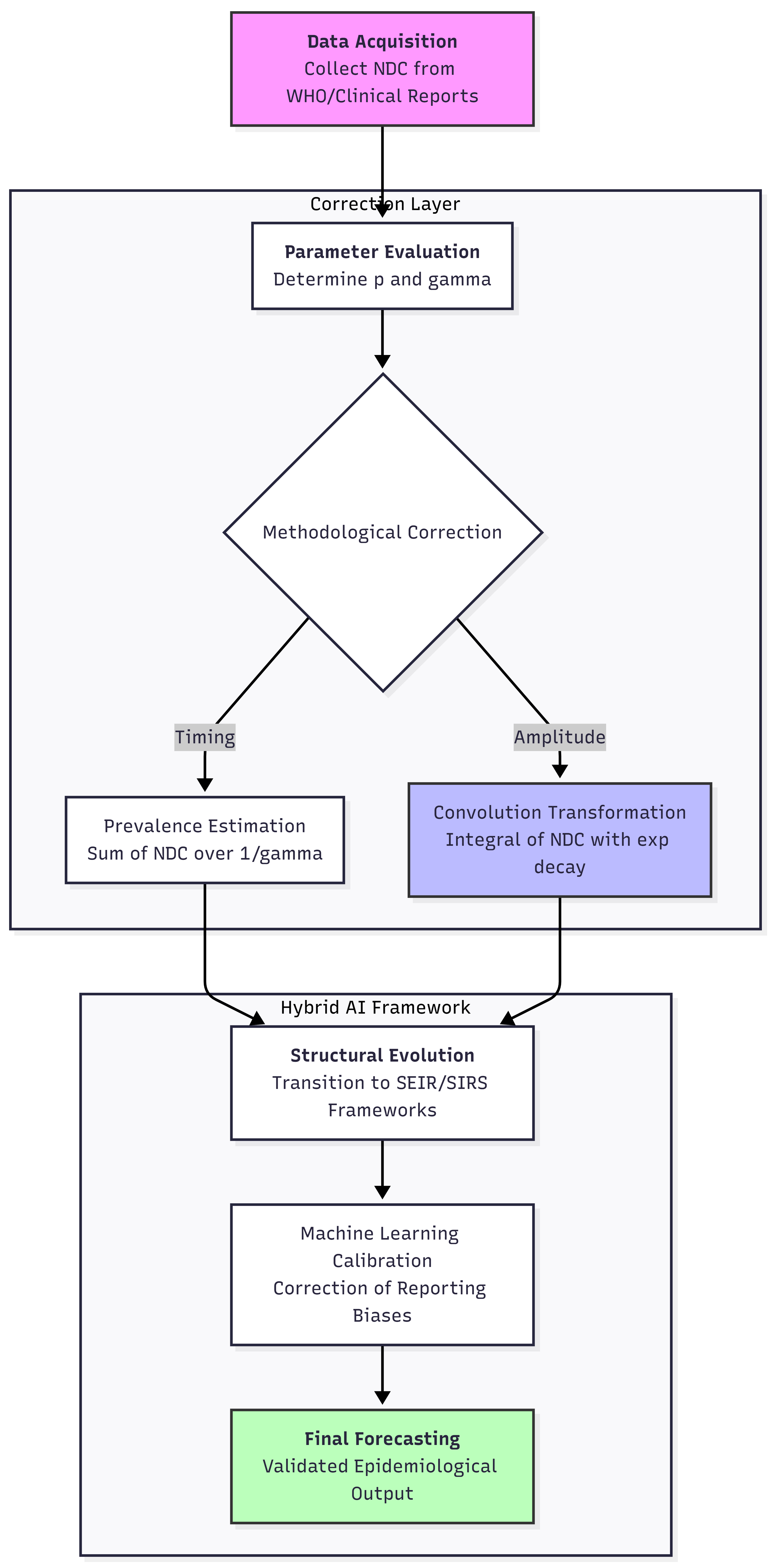}
    \caption{Workflow for practitioners to implement corrected SIR modeling. The process emphasizes the transition from New Daily Cases (Incidence) to Active Cases (Prevalence) through convolution.}
    \label{workflow}
\end{figure}
\section{ Solution for all models as SIR, SEIR or SIRS}

The errors done using $NDC(t)$ in place of the $I(t)$ usually done in $SIR$ model  happen also in $SEIR$ and $SIRS$ models. While these more complex models add biological realism (like latency or waning immunity), they still require active case counts (prevalence) for calibration, yet practitioners continue to feed them with raw daily reports (incidence). 

Practitioners often assume that moving from SIR to SEIR or SIRS automatically solves accuracy issues. However, as the table (\ref{tab:model_comparison}) shows,the errors on data misalignements will propagate to $SEIR$ and $SIRS$ parameters, generating errors in forecasting.


In these models, the errors manifest in the following ways:
\begin{itemize}
    \item SEIR (Latency): Even with an Exposed (E) compartment, practitioners often attempt to identify the new clinical cases directly with the Infectious (I) compartment. This ignores the fact that $I(t)$ is a stock variable representing those who have finished incubation but have not yet recovered, leading to incorrect estimates of the latent period.
    \item SIRS (Waning Immunity): In models where individuals return to the Susceptible (S) pool, using raw incidence to calibrate the ``heavy tail" of the curve leads to massive inaccuracies, leading to incorrect estimates of  the rate of immunity loss.
\end{itemize}
\section{Conclusion}

 The fundamental methodological error which correspond to equate raw incidence with prevalence is not mitigated by increasing model complexity. While the transition to SEIR and SIRS frameworks adds biological realism through latency and waning immunity, these structures remain mathematically based on stock variables as prevalence $I(t)$. Consequently, if practitioners calibrate these advanced models using New Daily Cases ($NDC$) without transformation, the structural additions will inherit the errors that can be up to 50\% peak underestimation and temporal misalignment observed in basic SIR applications. We argue that the proposed exponentially weighted convolution must serve as a universal ``pre-processing layer'' needed to correctly calibrate the model. We illustrate how this universal pre-processor protocol should be implemented in any applications of the epidemiological models based on compartments in figure (\ref{workflow}).

\begin{acknowledgments}
We acknowledge financial support from SECIHTI and SNII (M\'exico).
\end{acknowledgments}
\end{document}